\documentclass[%
reprint,
superscriptaddress,
amsmath,amssymb,
prb,
]{revtex4-2}

\usepackage{multirow}
\usepackage{graphicx}
\usepackage{dcolumn}
\usepackage{bm}
\usepackage{hyperref}

\begin{document}


\title{Tri-Hexagonal charge order in kagome metal CsV$_{3}$Sb$_{5}$ revealed by $^{121}$Sb NQR}

\author{Chao Mu}
\affiliation{Beijing National Laboratory for Condensed Matter Physics and Institute of Physics, Chinese Academy of Sciences, Beijing 100190, China}
\affiliation{School of Physical Sciences, University of Chinese Academy of Sciences, Beijing 100190, China}

\author{Qiangwei Yin}
\affiliation{Department of Physics and Beijing Key Laboratory of Opto-electronic Functional Materials $\&$ Micro-nano Devices, Renmin University of China, Beijing 100872, China}

\author{Zhijun Tu}
\affiliation{Department of Physics and Beijing Key Laboratory of Opto-electronic Functional Materials $\&$ Micro-nano Devices, Renmin University of China, Beijing 100872, China}

\author{Chunsheng Gong}
\affiliation{Department of Physics and Beijing Key Laboratory of Opto-electronic Functional Materials $\&$ Micro-nano Devices, Renmin University of China, Beijing 100872, China}

\author{Ping Zheng}
\affiliation{Beijing National Laboratory for Condensed Matter Physics and Institute of Physics, Chinese Academy of Sciences, Beijing 100190, China}

\author{Hechang Lei}
\email{hlei@ruc.edu.cn}
\affiliation{Department of Physics and Beijing Key Laboratory of Opto-electronic Functional Materials $\&$ Micro-nano Devices, Renmin University of China, Beijing 100872, China}

\author{Zheng Li}
\email{lizheng@iphy.ac.cn}
\affiliation{Beijing National Laboratory for Condensed Matter Physics and Institute of Physics, Chinese Academy of Sciences, Beijing 100190, China}
\affiliation{School of Physical Sciences, University of Chinese Academy of Sciences, Beijing 100190, China}

\author{Jianlin Luo}
\affiliation{Beijing National Laboratory for Condensed Matter Physics and Institute of Physics, Chinese Academy of Sciences, Beijing 100190, China}
\affiliation{School of Physical Sciences, University of Chinese Academy of Sciences, Beijing 100190, China}
\affiliation{Songshan Lake Materials Laboratory, Dongguan 523808, China}


\begin{abstract}
We report $^{121}$Sb nuclear quadrupole resonance (NQR) measurements on kagome superconductor CsV$_3$Sb$_5$ with $T_{\rm c}=2.5$ K. $^{121}$Sb NQR spectra split after a charge density wave (CDW) transition at $94$ K, which demonstrates a commensurate CDW state. The coexistence of the high temperature phase and the CDW phase between $91$ K and $94$ K manifests that it is a first order phase transition. The CDW order exhibits Tri-Hexagonal deformation with a lateral shift between the adjacent kagome layers, which is consistent with $2 \times 2 \times 2$ superlattice modulation. The superconducting state coexists with CDW order and shows a conventional s-wave behavior in the bulk state.
\begin{description}
\item[PACS numbers] 74.25.nj, 71.45.Lr, 76.60.Gv, 76.60.-k
\end{description}
\end{abstract}


\maketitle


The newly discovered superconductor $A$V$_3$Sb$_5$ ($A$ = K, Rb, Cs) possesses a quasi-two-dimensional kagome structure, which provides a platform to investigate the interplay of topology, electron correlation effects and superconductivity\cite{Ortiz2019New,Ortiz2020Cs,Ortiz2021KVSb,Lei2021Rb,jiang2021kagome}. They undergo a charge density wave (CDW) transition at $T_{\rm CDW}$= $78$ K, $103$ K, $94$ K and a superconducting transition at $T_{\rm c}$=$0.93$ K, $0.92$ K, and $2.5$ K for $A$ = K, Rb, Cs, respectively.
The superconducting state is found to be spin singlet with s-wave pairing symmetry in the bulk state\cite{Mu2021CPL, xu2021multiband, duan2021nodeless}. STM studies observed possible Majorana modes\cite{liang2021threedimensional} and pair density wave\cite{chen2021roton} in the surface state.
There are also experimentally observed residual density of states in the superconducting state\cite{zhao2021nodal}, which may be due to the competition between the superconductivity and the CDW\cite{chen2021double,Yu2021Unusual,du2021pressureinduced,zhang2021pressureinduced,chenXL2021highly,zhao2021nodal}.

\begin{figure}
\includegraphics[width=0.45\textwidth,clip]{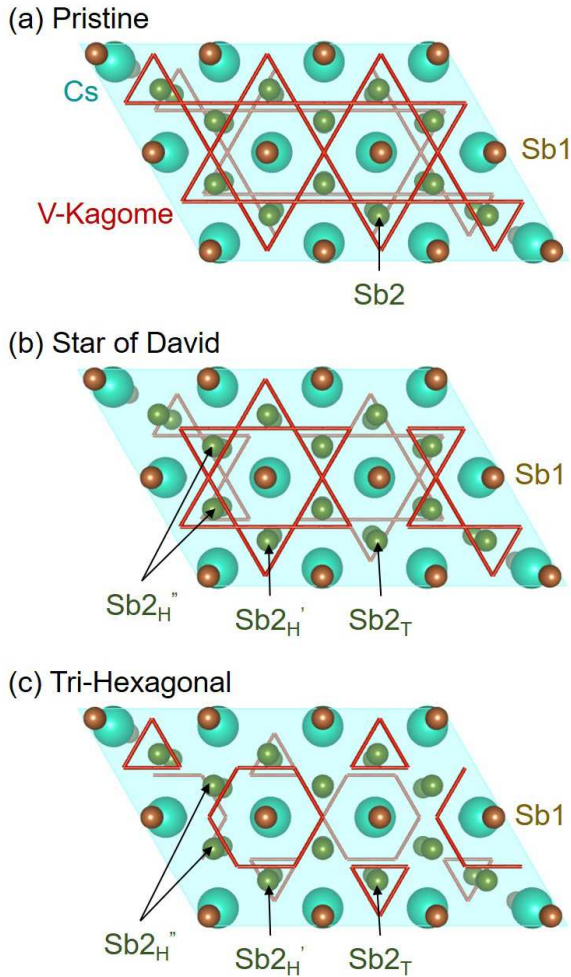}
\caption{\label{fig:Structure} Perspective view of crystal structures of CsV$_{3}$Sb$_{5}$ in (a) the pristine phase, (b) the star of David phase and (c) the Tri-hexagonal phase. There is a shift between adjacent layer in the star of David configuration and the Tri-hexagonal configuration.}
\end{figure}

The CDW order breaks time-reversal symmetry\cite{jiang2020discovery,Shumiya2021PRB,mielke2021timereversal} and leads to the anomalous Hall effect in the absence of magnetic local moments\cite{Yang2020Hall,Kenney2021muon,yu2021concurrence,feng2021chiral}.
Inelastic x-ray scattering and Raman scattering exclude strong electron-phonon coupling driven CDW\cite{li2021observation}, while optical spectroscopy supports CDW is driven by nesting of Fermi surfaces\cite{zhou2021origin}.
STM experiment found the CDW peak intensities at $3Q$ which breaks six-fold rotational symmetry\cite{Jiang2021NM}.
An additional unidirectional $1 \times 4$ superlattice is observed on the surface\cite{zhao2021cascade, chen2021roton, Liang2021PRX, wang2021electronic}, but is absent in the bulk state\cite{Li2021Spatial}.
The ``Star of David'' (SoD) and ``Tri-Hexagonal'' (TrH) structure configurations are proposed to be the likely candidates for CDW
structures\cite{ortiz2021fermi, Tan2021TrH}, as illustrated in Fig.\ref{fig:Structure}(b) and (c). SoD or TrH with a lateral shift between the adjacent Kagome layer results in a $2 \times 2 \times 2$ structure modulation\cite{Miao2021geometry}.
Density functional theory calculations show that the TrH deformation is preferred\cite{Ratcliff2021coherent, Tan2021TrH, ortiz2021fermi}. However, there is still lack experimental evidence whether SoD or TrH deformation is the ground state configuration in the CDW state.

In this work, we report nuclear quadrupole resonance (NQR) investigations on CsV$_{3}$Sb$_{5}$. The splitting of $^{121}$Sb spectra demonstrates that a commensurate CDW order forms at $94$ K with a first-order transition. Our results indicate that the charge order has a Tri-Hexagonal deformation which is $2 \times 2$ period modulation. The shift between the adjacent kagome layers induces a modulation in the $c$-direction and the CDW is a three-dimensional modulation with $2 \times 2 \times 2$ period. Spin-lattice relaxation rate measured at the peak of the CDW state shows a coherence peak just below $T_{\rm c}$, indicating that the superconducting state coexists with the CDW order and shows a conventional s-wave behavior.

Single crystals of CsV$_{3}$Sb$_{5}$ were synthesized using the self-flux method\cite{Mu2021CPL}. Superconductivity with $T_{\rm c} =2.5$ K is confirmed by dc magnetization measured using a superconducting quantum interference device (SQUID).
The NQR measurements were performed using a phase coherent spectrometer. The spectra were obtained by frequency step and sum method which sums the Fourier transformed spectra at a series of frequencies\cite{Clark1995FSS}. The spin-lattice relaxation time $T_{1}$ was measured using a single saturation pulse. Crystal structures were visualized in VESTA\cite{VESTA2011}.


\begin{figure*}
\includegraphics[width=0.98\textwidth,clip]{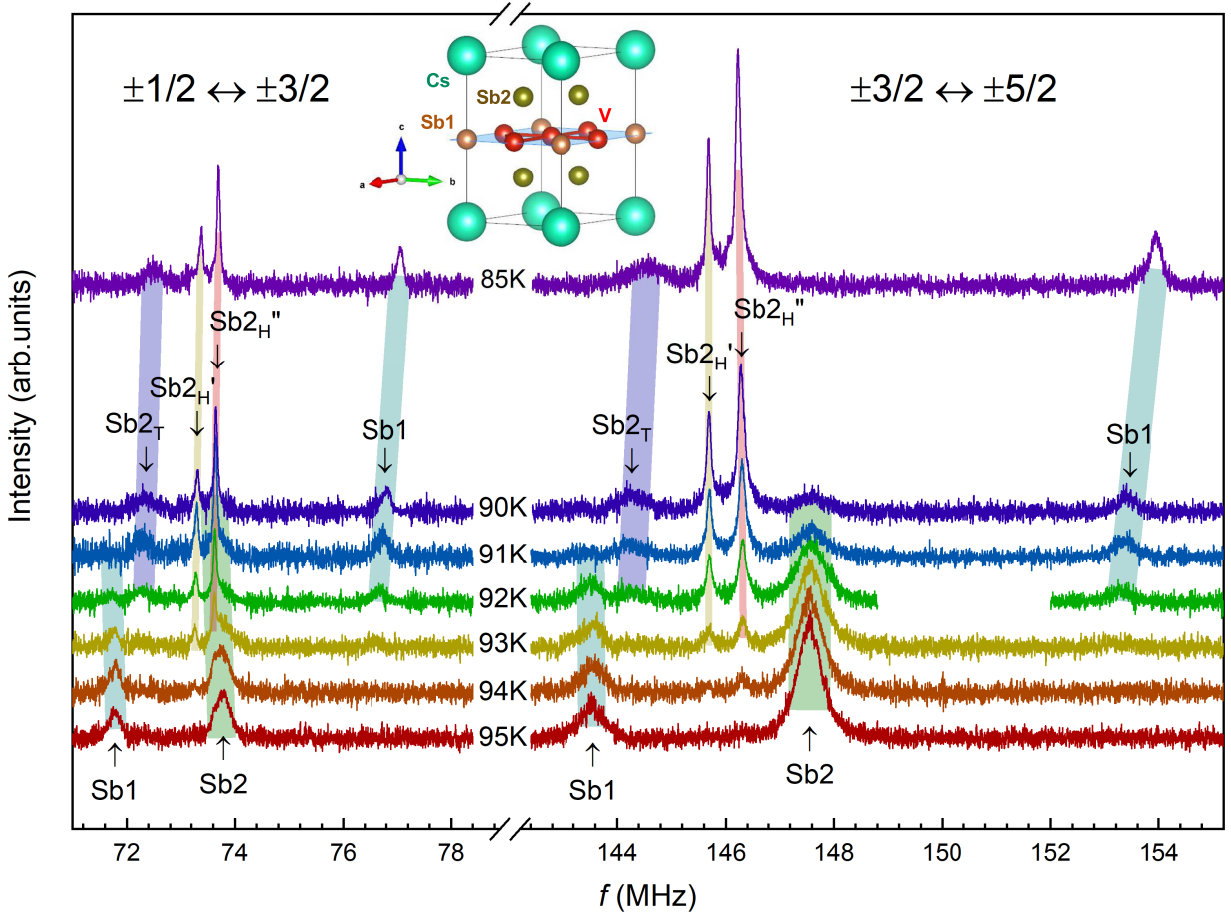}
\caption{\label{fig:SbSpectra} $^{121}$Sb-NQR spectra around $T_{\rm CDW}$. The peaks in the low frequency range are from $\pm\frac{1}{2} \leftrightarrow \pm\frac{3}{2}$ transition and the peaks in the high frequency range are from $\pm\frac{3}{2} \leftrightarrow \pm\frac{5}{2}$ transition. There are $2$ sites of Sb marked by Sb$1$ and Sb$2$ above $T_{\rm CDW}$. Both peaks of Sb$2$ split due to non-equivalent sites emerge below $T_{\rm CDW}$. The notations of Sb$1$, Sb$2_{\rm H}'$, Sb$2_{\rm H}''$ and Sb$2_{\rm T}$ are consistent with Fig. \ref{fig:Structure}(c). The spectra of all temperatures are presented in the supplemental material.}
\end{figure*}

Figure \ref{fig:SbSpectra} shows the NQR spectra of $^{121}$Sb in zero field. There are two different crystallographic sites of Sb in CsV$_{3}$Sb$_{5}$ with the atomic ratio of Sb$1:$ Sb$2=1:4$. Sb$1$ is located in the V-kagome plane, and Sb$2$ forms a graphite-like network between V-kagome layers, as shown in the inset of Fig. \ref{fig:SbSpectra} and Fig. \ref{fig:Structure} (a). It is easy to distinguish Sb$2$ peaks from Sb$1$ peaks by the spectral intensity.
The nuclear spin Hamiltonian of the interaction between quadrupole moment $Q$ and the electric field gradient (EFG) is described as
\begin{equation} \label{eq:vQ} 
\begin{aligned}
&\mathcal{H}_{\rm Q}
= &\frac{h\nu_{\rm Q}}{6}\left[ (3I_{z}^{2}-I^{2})+\frac{\eta}{2} (I_{+}^{2}+I_{-}^{2})  \right]
\end{aligned}
\end{equation}
where $\nu _{\rm Q}$ is the quadrupole resonance frequency along the principal axis ($c$-axis), $\nu _{\rm Q} \equiv \frac{3e^{2}qQ}{2I(2I-1)}$, with $eq=V_{zz}$. $\eta$ is an asymmetry parameter of the EFG, $\eta=\frac{V_{xx}-V_{yy}}{V_{zz}}$, where $V_{xx}$, $V_{yy}$, $V_{zz}$ are the EFGs along the $x$, $y$, $z$ directions respectively.
$^{121}$Sb with $I=5/2$ has two NQR peaks for each site, $|\pm\frac{1}{2}\rangle \leftrightarrow |\pm\frac{3}{2}\rangle$ transition at low frequency and $|\pm\frac{3}{2}\rangle \leftrightarrow |\pm\frac{5}{2}\rangle$ transition at high frequency.

Four NQR transition peaks can be clearly identified above $95$ K.
Below $T_{\rm CDW}=94$ K, the intensity of original peaks decreases gradually and four sets of new peaks emerge, indicating that the CDW transition is commensurate and four unequal sites of Sb atoms appear in the CDW state. The coexistence of two phases between $91$ K and $94$ K demonstrates that the CDW transition is of first-order due to a simultaneous superlattice transition\cite{Ortiz2020Cs}. It is consistent with NMR studies on $^{51}$V which also show coexistent behavior\cite{Mu2021CPL, song2021orbital, luo2021starofdavid}.
The peaks below $T_{\rm CDW}$ are complex and cannot be distinguished by intensity alone.

\begin{figure}[htb]
\includegraphics[width=0.49\textwidth,clip]{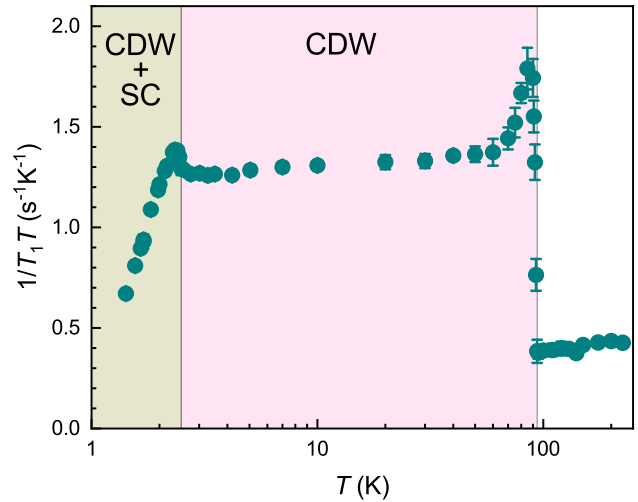}
\caption{\label{fig:SbT1T} Temperature dependence of $^{121}(1/T_{1}T)$ of Sb$1$ site. $1/T_{1}T$ jumps at $T_{\rm CDW}$. A Hebel-Slichter coherence peak appears just below $T_{\rm c}$.}
\end{figure}

Two structure configurations named as ``Star of David'' and ``Tri-Hexagonal'' have been proposed in the CDW state\cite{ortiz2021fermi,Ratcliff2021coherent,Tan2021TrH}. We can assign the Sb$2_{\rm H}$ and Sb$2_{\rm T}$ sites by their symmetry.
In SoD configuration, as shown in Fig. \ref{fig:Structure} (b), Sb$2_{\rm H}$ is inside the star and Sb$2_{\rm T}$ is outside the star.
In TrH configuration, as shown in Fig. \ref{fig:Structure} (c), Sb$2_{\rm T}$ is located at the center of the trigonal V-network and Sb$2_{\rm H}$ is outside.
A lateral shift between the adjacent kagome layer form modulation on the $c-$axis\cite{Miao2021geometry,Ratcliff2021coherent}, which result in a nematic state with only $C_{2}$ rotation symmetry and make Sb$2_{\rm H}'$ site different from Sb$2_{\rm H}''$ site\cite{Yuli2021}.
The subscript of ``H'' denotes the sites around the hexagon and the subscript of ``T'' denotes the sites at the triangle center. The four new peaks in Fig. \ref{fig:SbSpectra} correspond to these four sites, Sb$1$, Sb$2_{\rm H}'$, Sb$2_{\rm H}''$ and Sb$2_{\rm T}$.
The spin-lattice relaxation time, $T_{1}$, is employed to track the evolution of the peaks. $1/T_{1}T$ of the peak around $77$ MHz evolutes from Sb$1$, as shown in Fig. \ref{fig:SbT1T}. $1/T_{1}T$ of the peaks around $74$ MHz marked by Sb$2_{\rm H}'$ and Sb$2_{\rm H}''$ are same and evolute from Sb$2$, see the Fig. $2$ in Ref \cite{Mu2021CPL}. The peak around $72$ MHz has weak intensity and $T_{1}$ measurement has large uncertainty just below $T_{\rm CDW}$. Fortunately, there is only Sb$2_{\rm T}$ site left, so the peak around $72$ MHz is attributed to it.
In addition, the asymmetry parameters $\eta$ can also be used to distinguish them.

\begin{figure*}
\includegraphics[width=0.98\textwidth,clip]{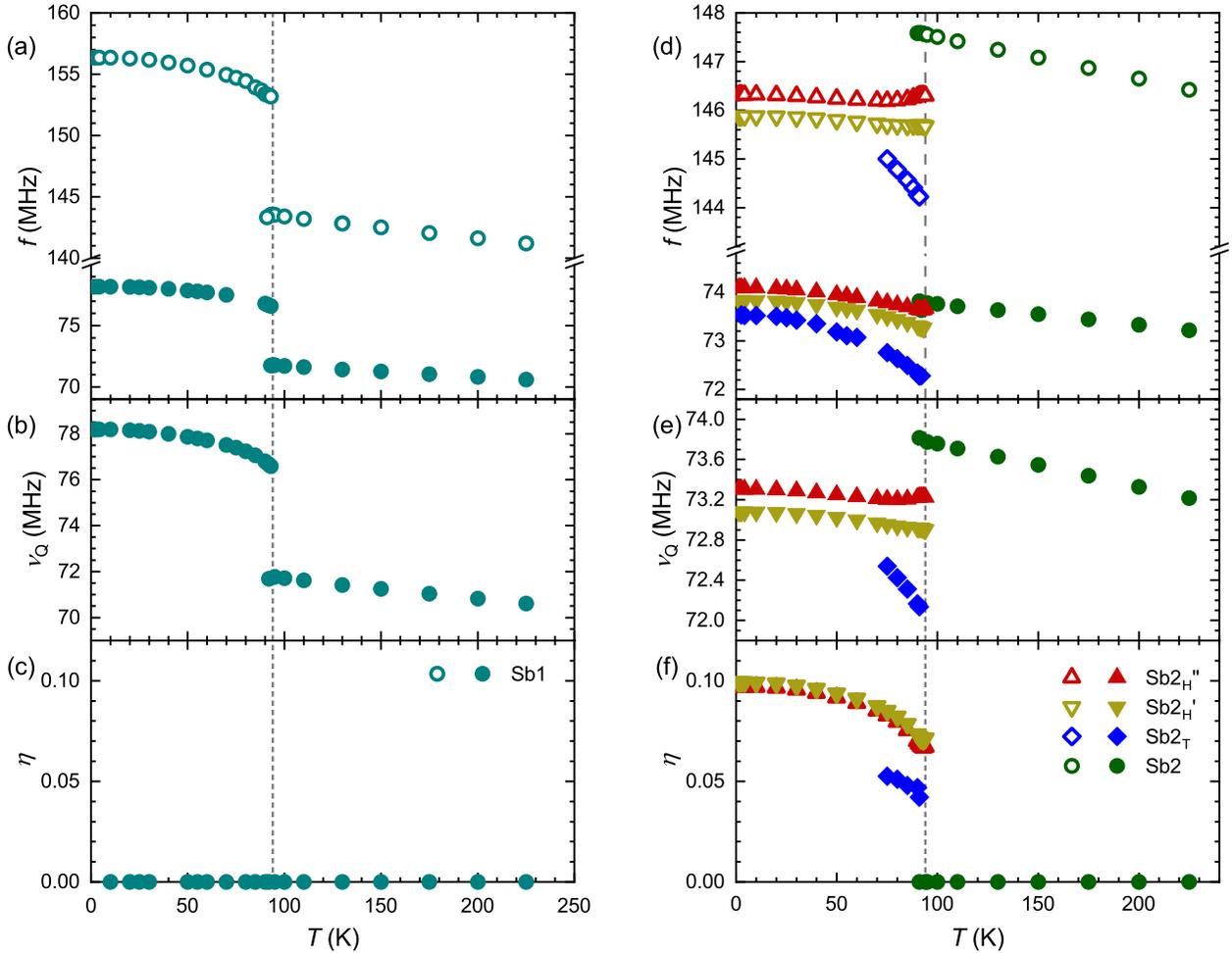}
\caption{\label{fig:Sbfvn} Temperature dependence of (a)(d) peak positions, (b)(e) $\nu _{\rm Q}$ and (f)(g) $\eta$ of Sb$1$ site and Sb$2$ site, respectively. The vertical dashed lines indicate the transition temperature of CDW.}
\end{figure*}

All peak positions are summarized in Fig. \ref{fig:Sbfvn} (a)(d). Combining the frequencies of $|\pm\frac{1}{2}\rangle \leftrightarrow |\pm\frac{3}{2}\rangle$ transition and $|\pm\frac{3}{2}\rangle \leftrightarrow |\pm\frac{5}{2}\rangle$ transition, $\nu _{\rm Q}$ and $\eta$ are calculated, as shown in Fig. \ref{fig:Sbfvn}(b)(c)(e)(f). The $\eta$ at all sites is zero above $T_{\rm CDW}$, reflecting the isotropy of EFG in the $xy$-plane. Below $T_{\rm CDW}$, $\eta$ of Sb$2$ changes to finite values, indicating the in-plane symmetry is broken and unequal sites of Sb$2$ are generated. The slightly decreasing $\nu _{\rm Q}$ implies that the main change is not the EFG strength, but the EFG direction at Sb$2$ sites. On the other hand, Sb$1$ in kagome plane has high symmetry, so the EFG at Sb$1$ site changes frequency obviously without asymmetry. When the electric field modulation period is twice the lattice constant, the highly symmetrical sites are not affected by the modulation\cite{Li2016TaPdTe}, so Sb$1$ peak does not split like Sb$2$ peak.

Next, we will distinguish between the two configurations of TrH and SoD. The CDW has a three-dimensional modulation and the modulation in $c$-axis will make Sb$2$ plane has different pattern from the kagome plane.
Therefore, the different between TrH and SoD in the kagome plane should be detected by Sb$1$ rather than Sb$2$.
In TrH deformation, V atoms are closer to Sb$1_{\rm}$ site, so $\nu _{\rm Q}$ of Sb$1_{\rm}$ moves a lot toward high frequency, as shown in Fig. \ref{fig:Sbfvn} (b). In SoD deformation, on the other hand, V atoms move away from Sb$1_{\rm}$ site, so $\nu _{\rm Q}$ of Sb$1_{\rm}$ should not increase, which is inconsistent with the NQR spectra.
The Sb$2_{\rm H}$ peaks split to Sb$2_{\rm H}'$ and Sb$2_{\rm H}''$ peaks due to a lateral shift between adjacent layer in the Tri-hexagonal configuration. The intensity ratio of two peaks are $1 : 2$, which is consistent with the atoms ratio. NQR frequency is determined by EFG, so the NQR spectra indicates that the electronic modulation pattern is formed simultaneously with the structural distortion. Our results favor TrH deformation over SoD deformation and are consistent with $2a \times 2a \times 2c$ pattern. If $4a$ pattern exists, it should only exist on the surface\cite{Tan2021TrH, Li2021Spatial}.

$\nu _{\rm Q}$ and $\eta$ jump at $T_{\rm CDW}$ and then change gradually in the CDW state. They remain constant when entering the superconducting state, indicating the coexistence of superconductivity and the CDW order. $^{121}(1/T_{1}T)$ measured at the peak in the CDW order state shows superconducting transition, as shown in Fig. \ref{fig:SbT1T}, confirming the coexistence. $^{121}(1/T_{1}T)$ shows a clear Hebel--Slichter coherence peak just below $T_{\rm c}$ and then rapidly decreases at low temperatures, as shown in Fig. \ref{fig:SbT1T}, indicating that CsV$_{3}$Sb$_{5}$ is a conventional s-wave superconductor. It is consistent with our previous result measured at Sb$2$ site\cite{Mu2021CPL}.

In conclusion, we have performed $^{121}$Sb NQR studies on CsV$_{3}$Sb$_{5}$. The variation of the NQR spectra indicates a first-order commensurate CDW transition occurs at $94$ K. We identified four Sb sites in the CDW state which give restriction on the CDW order. The charge order can be understood in terms of Tri-Hexagonal deformation with lateral shift between the adjacent kagome layers, which has a $2 \times 2 \times 2$ period. In the superconducitng state, $^{121}(1/T_{1}T)$ of Sb$1$ site in kagome planes shows conventional s-wave behavior, just as $^{121}(1/T_{1}T)$ of Sb$2$ site between kagome planes. The superconducting phase coexist with CDW order in the bulk state.

This work was supported by the National Key Research and Development Program of China (Grant No. 2017YFA0302901, 2018YFA0305702, 2018YFE0202600, and 2016YFA0300504), the National Natural Science Foundation of China (Grant No. 12134018, 11921004, 11822412 and 11774423), the Beijing Natural Science Foundation (Grant No. Z200005), and the Strategic Priority Research Program and Key Research Program of Frontier Sciences of the Chinese Academy of Sciences (Grant No. XDB33010100).



%

\end{document}